# The quiet chromosphere

## Old wisdom, new insights, future needs

Robert J. Rutten

1 Sterrekundig Instituut, Utrecht University, Utrecht, The Netherlands
2 Institute of Theoretical Astrophysics, University of Oslo, Oslo, Norway
  e-mail: R.J.Rutten@uu.nl

**Abstract.** The introduction to this review summarizes chromosphere observation in two figures. The first part showcases the historical emphasis on the eclipse chromosphere in the development of NLTE line formation theory and criticizes 1D modeling. The second part advertises recent breakthroughs after many decades of standstill. The third part discusses what may or should come next.

## 1. Introduction

Figure 1 summarizes how the chromosphere looks. The caption summarizes why it is of interest. Both come from political documents. Figure 2 inventorizes the optical chromospheric diagnostics.

Observationally, the finely structured, highly dynamic fibrilar nature of the chromosphere requires $0.1''$ angular resolution, 1 s time resolution, spectral resolution $2 \times 10^5$ across the few optical spectral lines that sample the chromosphere (Fig. 2), and $10^{-5}$ sensitivity in their Stokes parameters to measure magnetic fields. New vistas realizing such data collection are opened by the advent of open-telescope technology (initiated with the DOT), of adaptive optics (initiated at the DST, SST, and VTT), and of numerical post-processing (speckle and MOMFBD reconstruction). These three technological advances together enable effective use of larger aperture than the present generation of 0.5–1 m solar telescopes, and larger aperture is indeed required to reach the numbers above. The chromosphere has so become the principal science driver for the large optical solar telescopes being built (American ATST) and under consideration (Indian NLST, European EST, Japanese SOLAR-C/plan-B).

Interpretationally, the chromosphere requires numerical MHD simulation far beyond schematic analytic or cartoon physics. It must include 3D radiative transfer, scattering, partial redistribution for some lines (certainly for Mg II h & k and Ly$\alpha$), non-equilibrium ionization (certainly for hydrogen), and possibly multi-fluid description. Spatial resolution and spatial extent are obvious desires. These challenges are formidable and make the chromosphere the frontier in forward simulation efforts. Approaches trying inversion remain premature.

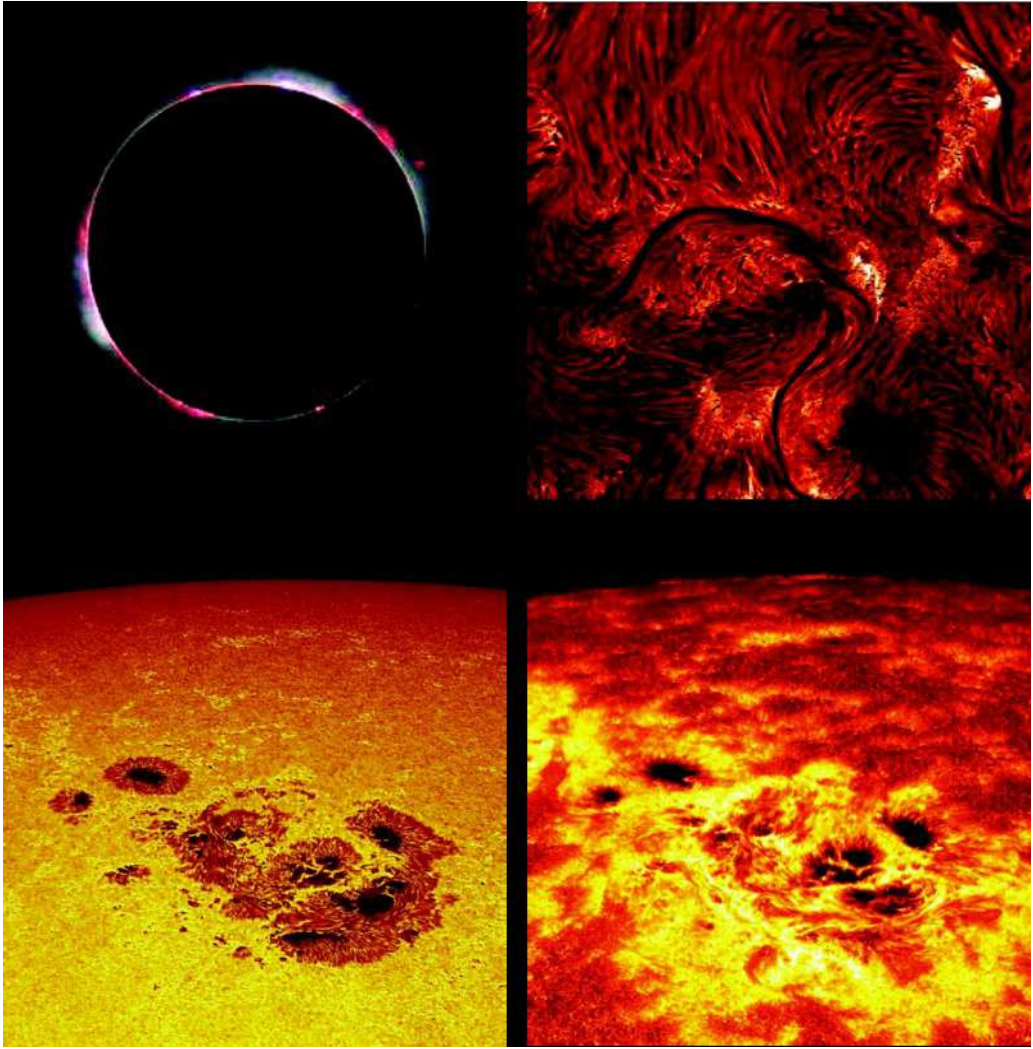

**Fig. 1.** The solar chromosphere. This image assembly opens the *"White Paper on the solar chromosphere"* for the US Astro2010 Decadal Survey by Ayres et al. (2009). The eclipse image illustrates the naming, by Lockyer (1868), of the chromosphere as an uneven but ubiquitous, colorful solar envelope radiating in the prominence lines (Fig. 2). The DOT Hα filtergram illustrates that the chromosphere consists of a mass of slender fibrils that chart magnetic-field connectivity over extended areas, with finer-grained structure in active regions. The bottom images, also from the DOT and also F. Snik-colored, illustrate the large change in active-region appearance from the photospheric surface (G band, left) to the low chromosphere (Ca II H, right).

The enigmatic, dynamic, phenomena-rich chromosphere is the transition between the solar surface and the eruptive outer solar atmosphere. At and below the surface, the gas pressure dominates over the magnetic pressure outside sunspots and compresses magnetic fields into slender, upright fluxtubes that appear as tiny white dots in the G-band image. They expand into space-filling field in the chromosphere, above which magnetic forces reign in setting the coronal structure and dynamics. The chromosphere harbors and constrains the mass and energy loading that cause filament eruptions and flares, together governing near-Earth space weather.

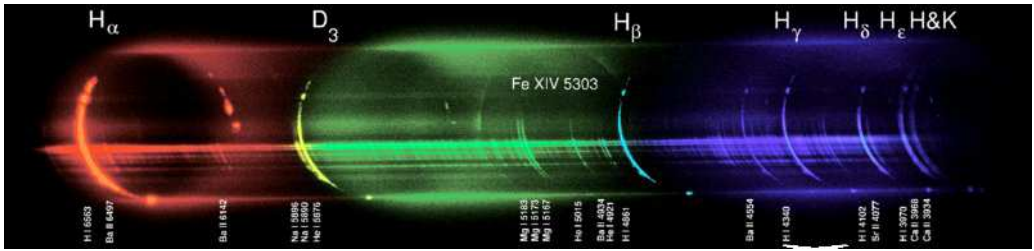

**Fig. 2.** Flash spectrum taken by Manfred Rudolf of the EurAstro Team Szeged I on August 11, 1999 (`http://www.eurastro.de/webpages/MRSPECT.HTM`). He used a 200 grooves/mm blazed grating in front of a 500 mm lens mounted on a classical 35-mm camera with Ektachrome 200 slide film. The chromaticity of the lens caused defocus at both ends; the chromaticity of the film influences the colors. A much sharper monochrome flash spectrum covering about the same wavelength region is shown in Plate 3 of Dunn et al. (1968). The Fe XIV 5303 Å green coronal line is seen as a vague outline of the dark moon.

The stronger lines together constitute the purple color of the off-limb chromosphere in the first panel of Fig. 1. In red-only imaging the Hα color whitens into pink by Thomson scattering (Jejčič & Heinzel 2009).

In 1986 Lockyer had not yet observed during totality, but using his new low-dispersion prominence spectrohelioscope outside eclipse he found that *"the prominences are merely local aggregations of a gaseous medium which entirely envelopes the sun. The term* Chromosphere *is suggested for this envelope, in order to distinguish it from the cool absorbing atmosphere on the one hand, and from the white light-giving photosphere on the other. The possibility of variations in the thickness of this envelope is suggested [ . . . ] Under proper instrumental and atmospheric conditions, the spectrum of the chromosphere is always visible in every part of the sun's periphery [ . . . ] Two of the lines correspond to Fraunhofer's C and F; another lies 8° or 9° (of Kirchhoff's scale) from D towards E. There is another bright line, which occasionally makes its appearance near C, but slightly less refrangible than that line. It is remarked that the line near D has no corresponding line ordinarily visible in the solar spectrum."*, where C is Hα, F is Hβ, the non-Fraunhofer line near D is He I D₃, and the less refrangible line near Hα is Ba II 6497 Å.

Adherence to Lockyer's definition of chromosphere implies that only the stronger lines seen above, plus the Ca II infrared lines and He I 10830 Å outside this range, are chromospheric. Optical diagnosis of the chromosphere is therefore limited to Balmer lines, He I lines, Ca II lines, and perhaps Ba II lines. The Na I D lines are barely visible here and are indeed not chromospheric in the simulation of Leenaarts et al. (2010). Hα stands out as the principal chromospheric line, although H & K are much stronger in the photospheric spectrum conform Saha-Boltzmann partitioning. The chromospheric prominence of Hα is probably due to enormous overpopulation resulting from frequent ionization by shocks followed by cascade recombination and slow equilibrium settling.

The dominance of Hα in the flash spectrum implies that the ensemble of structures that together constitute the projected forest of spicules (of whatever type) at the limb away from prominences reaches its largest collective optical thickness (not "depth", please!) along the tangential line of sight in Hα. In radial viewing the extended fibril canopies displayed in the second panel of Fig. 1 are therefore probably also more opaque in Hα than in any of the other lines. These fibril canopies contribute much opacity to the highly confused lower parts of the off-limb spicule forest. Dynamic fibrils may reach higher into less confusion further out from the limb, and likely represent much of the classical spicules. Spicules-II reach yet higher but are too slender and dynamic to have been resolved in classic seeing-hampered long-exposure H & K imaging. Their on-disk straw/RBE counterparts are far from obvious in Fig. 1, perhaps appearing as diffuse brightness haze above near-limb network in the fourth panel.

- R. N. Thomas & R. G. Athay, 1961, *"Physics of the Solar Chromosphere"*
- J. M. Beckers, 1964, *"A Study of the Fine Structures in the Solar Chromosphere"*, PhD Thesis Utrecht University
- A. M. Title, 1966, *"A Study of Velocity Fields in the Hα Chromosphere by Means of Time-Lapse Doppler Movies"*, PhD Thesis CalTech
- C. de Jager (Ed.), 1968, *"The structure of the quiet photosphere and the low chromosphere"*, Procs. Bilderberg Conference
- R. J. Bray and R. E. Loughhead, 1974, *"The Solar Chromosphere"*
- R. G. Athay, 1976, *"The Solar Chromosphere and Corona: Quiet Sun"*
- B. W. Lites (Ed.), 1985, *"Chromospheric Diagnostics and Modeling"*, Procs. NSO Summer Conference
- P. Ulmschneider, E. R. Priest & R. Rosner (Eds.), 1991, *"Mechanisms of Chromospheric and Coronal Heating"*, Procs. Int. Heidelberg Conference
- M. Carlsson (Ed.), 1994, *"Chromospheric Dynamics"*, Procs. Oslo Mini-Workshop
- P. Heinzel, I. Dorotovič, R.J. Rutten (Eds.), 2007, *"The Physics of Chromospheric Plasmas"*, Procs. Coimbra Conference

**Table 1.** All books with titles containing "chromosphere" in my bibtex file books.bib.

## 2. Old wisdom

### 2.1. Books

Table 1 lists wisdom. Half are monographs, half conference proceedings. The copy of the first (Thomas & Athay 1961) in the NSO/SP library is adorned with the comments shown in Fig. 3. They made me read De Jager's critique, not listed on ADS under his name but as 1962ZA.....55...66T and 1962ZA.....55...70W. Five pages in solid German[1]. The first sentences in translation:

*"The title of this book promises more than the content offers. Anyone who has ever seen the enchanting structure of the chromospheric surface through an Hα filter or spectrohelioscope or has observed the solar limb will immediately – upon hearing the title "Physics of*

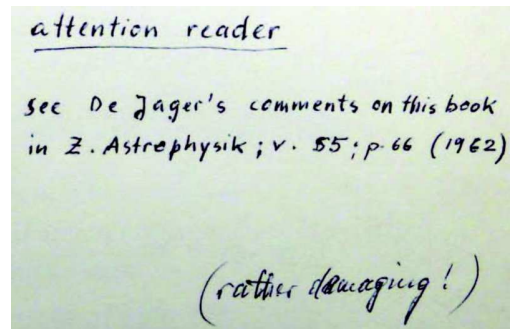

**Fig. 3.** Comments on the title page of Thomas & Athay (1961) in the Sunspot library.

*the Chromosphere" – expect explanation of the dynamics of these gases. He will think of the problems posed by sound, shock, and gravity waves and of their energy dissipation. He may wonder what the authors think of the role of magnetic fields and MHD waves, and to what extent these set the various structures of quiet and active regions in this remarkable part of the Sun. But he won't find any of these in this book: such problems are barely mentioned. The book is limited almost exclusively to spectroscopic issues."*

---

[1] During WW-II C. de Jager escaped the German Arbeitseinsatz by hiding over two years in Utrecht's Sterrewacht Sonnenborgh. Before that, German was the principal language of continental astrophysics. The last vestige was Unsöld's monumental *"Physik der Sternatmosphären"* (1955) which you probably haven't read. H. Panofsky and K. Pierce translated it into English but got no permission for publication. There are a few mimeograph copies at NSO Tucson. Maybe one should be scanned for ADS?

## 2.2. Limb chromosphere

The chromospheric physics in Thomas & Athay (1961) is indeed largely limited to NLTE line formation in the flash spectrum observed by the HAO expedition to the 1952 Sudan eclipse. Thanks to K. Reardon and the NSO/SP library, I belatedly found that similarly monumental NLTE flash-spectrum analysis was given by Menzel (1931). Osterbrock (2002) summarized it as:

*"Menzels Magnum Opus [. . . ], a huge, 303-page, quarto-size volume on the physical nature of the solar chromosphere. Roughly 200 pages are tables. [. . . ] The rest of the paper contains Menzels derivations of the curve of growth for emission lines with continuous absorption included, the ionization equations in usable form, and the theoretically expected gradients (scale heights). It is a masterful paper, and the beginning of the true scientific study of the chromosphere. [. . . ] But Aitken and Wright were highly skeptical of Menzels unexpected results on high temperatures and large deviations from thermodynamic equilibrium in the chromosphere. To most astronomers of that time it seemed unreasonable that the temperature in the outermost layers of the Sun could be higher than at its surface. Their training, experience, and knowledge left them unable to understand and evaluate the straightforward observational evidence Menzel had to back up his statements."*

At the time, the review by Stewart (1932) indeed shied at the math:

*"The central purpose of Menzel's study has been shifted toward the theoretical, a little too far perhaps. [. . . ] The subject is an exceedingly complex one – in its physical implications as well as in its astrophysical applications; and no completely adequate treatment is possible in the present state of theory and observation. The algebraic intricacies of certain phases of the subject have tempted more than one well-known investigator to develop his equations far beyond the point where the necessary physical and observational knowledge fails; and the resultant discussions must be in part invalid. [. . . ] Menzel attempts to keep relatively close to established physical principles; but it would have been difficult everywhere to make clear the distinction between observational fact and theoretical extrapolation."*

I cannot remember being told about Menzel's work while working on flash spectra for my own thesis. (Even now one wouldn't find it since it is not even listed on ADS.) I should confess that I also had a hard time with the math. My main guide became Jefferies' (1968)[2] revolutionary *Spectral Line Formation*, then came Mihalas' (1970) magisterial *Stellar Atmospheres*, and finally the comprehensive description by Vernazza et al. (1981) of the beautiful but 1D solar analogon VAL3C. These three plus C. Zwaan's excellent lecture notes[3] let me grasp the math, as laid down in my own lecture notes (Rutten 2003). In hindsight, it seems fair to think that the limb chromosphere contributed to the start of this mature understanding of NLTE line formation. The latter is needed to understand and diagnose the chromosphere, but De Jager (1962) was perfectly right in emphasizing waves and magnetism as the meat of chromospheric physics.

## 2.3. Disk chromosphere

The birefringent filter invented by Lyot (who died on the way back from the Sudan eclipse) made the disk chromosphere synonymous with the fine structure seen in Hα. The main inventory was made by Beckers (1964) in his thesis (the second book in Table 1). Beckers also wrote the book on off-limb chromosphere fine structure in the form of his spicule reviews (1968, 1972). Not much progress in explaining chromospheric fine structure followed during three decades. The general opinion was that

---

[2] The same year as Dunn et al. (1968), the comprehensive 1962 flash spectrum compilation for which Jefferies performed the laborious intensity and wavelength measurements. Like Menzel's line list it had almost no impact, signaling the end of the importance of eclipse observation. Most subsequent eclipse expeditions were fruitless.

[3] Neatly handwritten. The last version was latexed but remained in Dutch. E.H. Avrett's lecture notes, also neatly handwritten but in English, have now been latexed by him and are available at his website (Google Avrett CfA).

Hα images are most admirable but too complex to tackle.

What was tractable was the spatio-temporal mean, culminating in VAL3C. However, its underlying assumption that the fine structure is a small-amplitude fluctuation that may be meaningfully averaged was upset by the identification of Ca II H & K internetwork grains as acoustic shock interference by Carlsson & Stein (1992, 1994, 1997). The second reference is the next to last book in Table 1 and already contained Skartlien's plot, published later in Carlsson & Stein (1995), that shows how the non-linear Planck function sensitivity in the ultraviolet upsets such averaging when the fluctuations are large and can reproduce VAL3C as shocks superimposed on a radiative-equilibrium stratification. This happens in the heart of internetwork cells where the magnetic canopy, outlined by Hα fibrils, is high enough for the shocks to fully develop and interact. I termed this domain of acoustic domination "clapotisphere"[4] (Rutten 1995). Presumably, it is also full of internal gravity waves which remain hard to diagnose (Rutten & Krijger 2003; Straus et al. 2008).

While the VAL3C quiet-chromosphere description was upset by acoustic internetwork shocks, for plage and network the VAL approach of 1D multi-component description remains popular in solar irradiance studies. They account for network and plage brightening with models as those of Unruh et al. (1999) that postulate outward temperature divergence between magnetic concentrations and their less-magnetic environment. However, in photospheric diagnostics including the Mn I lines (Vitas et al. 2009) network and plage brightening stems from multi-D Spruit (1976) "hot-hole-in-the-surface" viewing. Towards the limb it stems from multi-D "hot-granule-interior" viewing. Such viewing requires at least 2D modeling, as Solanki's classical multi-ray wine glass fluxtubes (e.g.,

Bünte et al. 1993). Magnetic concentrations may so appear bright even when they and their environment obey non-heated radiative-equilibrium stratifications.

In Rutten (2010) I have advocated that such hole viewing may even dominate in Na I D$_1$ and wide-band Ca II H & K brightening, but the recent simulatory synthesis of Na I D$_1$ by Leenaarts et al. (2010) suggests that magnetic concentrations brighten in this line from the combination of downdraft Dopplershift and slight fluxtube heating around $h = 200$ km. The latter produces temperature stratifications like VAL3C but shifted down over the Wilson depression, with a temperature minimum near $h = 100$ km. The small temperature rise is attributed to Joule heating by Carlsson et al. (2010) in this volume. One might call it "chromospheric" in the VAL tradition of calling layers above a temperature minimum chromospheric, but this seems awkward for heights where the environment is mid-photosphere, far below the fibril chromosphere seen in Hα.

## 3. New insights

I think it is fair advice to a student embarking on chromospheric fine structure to read Beckers' thesis and spicule reviews and then skip the whole literature until the recent papers by De Pontieu and his collaborators, in particular the Oslo group. They usually combine superb MOMFBD-ed Hα and/or Ca II 8542 Å observations from the SST with simulations. Their first breakthrough was their analysis of "dynamic fibrils" (Hansteen et al. 2006; De Pontieu et al. 2007a), based on the re-realization by De Pontieu et al. (2004) that inclined fluxtubes have lower cutoff frequencies – as derived earlier by Michalitsanos (1973), Bel & Leroy (1977), and Suematsu (1990). In my opinion this was the first solid identification of a chromospheric fine-structure phenomenon, displaying specific Hα behavior, with a plausible physical mechanism, repetitive *p*-mode shock driving.

The next was the discovery of "spicules-II", not with the SST but with Hinode's only effective diagnostic of the chromosphere, off-limb Ca II H emissivity (De Pontieu et al.

---

[4] Taken from a kayaking manual. Much later, Wedemeyer-Böhm & Wöger (2008) introduced "fluctosphere" for a similar shock-dominated regime in CO5BOLD simulations which may or may not represent a valid solar analogon (see below for the may not).

2007c). This is even more exciting since these seem to display the major chromospheric action in quiet-sun coronal mass and energy loading as fast wavy outflows that likely arise from separatrix shear and component reconnection. The recent analyzes of De Pontieu et al. (2009) and McIntosh & De Pontieu (2009) connect them to the transition region and corona. They are hard to see on the disk[5]. As remarked by De Pontieu et al. (2007b), we had glimpsed them earlier (Rutten 2006, 2007) in a near-limb Ca II H DOT movie[6] as "straws", long, slender, highly dynamic features that emanate from network and are seen bright against the dark internetwork background whatever their Dopplershift through the wide-passband DOT Ca II H filter. They were then found on-disk as "rapid blue-shifted excursions" (RBEs) by Langangen et al. (2008) and described in more detail by Rouppe van der Voort et al. (2009) as very fast dark blobs at large blueshift in Ca II 8542 Å and Hα, ejected up and out, and hacked up in pieces per SST/CRISP tuning according to their accelerative Dopplershift.

The best way to observe these features is likely as straws, over their whole length whatever their segmental Dopplershift, with the wide-band filter imaging in Mg II h & k by IRIS because their h & k emissivity is much larger than for H & K while the coherently scattered internetwork h & k wing background is much darker than for H & K.

## 4. Future needs

**Instrumentation.** Blind deconvolution (MOMFBD, Van Noort et al. 2005) is replacing speckle reconstruction as the workhorse for numerical image restoration. It needs fewer frames and so permits faster cadence, and it delivers better quality when the seeing is good. Even at La Palma this is the case only frustratingly rarely. It is good that, in EST context, serious comparative seeing monitoring there and at Izaña is finally being realized. La Palma has its oft-disturbing Caldera, Izaña the wake-prone Teide. A large, cool lake filling a steep, high, dead, mid-ocean, low-latitude, solitary cone crater to its rim would be better – but none seems around.

Fabry-Pérot's have replaced Lyot filters as workhorse for high-resolution chromosphere diagnosis. They permit better profile sampling at faster cadence. It would be good to have them working also in space where the immersion oil used in Lyot-type filters gives bubble problems. So far, they only operate in the red where only the Ca II IR lines, Hα, and He I 10830 Å are chromospheric. It would be good to add Hβ, D₃, and Ca II H & K. IRIS will likely demonstrate that Mg II h & k provide yet better diagnostics even though they scatter most coherently (Owocki & Auer 1980).

Slit spectrometers are out since their slits limit the viewing to the wrong place at the wrong time and do not permit MOMFBD restoration. IRIS does not need the latter, being in space, and will use narrow-strip fast-scan sampling to make the best of the former. Full-field imaging spectroscopy would be better. Fiber and/or lenslet field reformatting remains hopeful (Rutten 1999; Lin et al. 2004). Imaging spectroscopy in the ultraviolet including Lyman lines would be ideal but seems a distant dream.

**Observation.** Lockyer's (1868) definition of the term chromosphere as prominence-spectrum envelope (Fig. 2) defines the Hα fibril canopy to be its on-disk manifestation. Wherever the Sun is at least slightly active it appears as a dense mass of optically thick, finely structured and highly dynamic cell-spanning fibrils overlying the internetwork clapotisphere. However, as striking as this fibril canopy is (called "enchanting" by De Jager 1962; "bewildering" suits too), the more interesting action seems to be in the up-and-out straws/RBEs. They are hard to see, require multi-line narrow-band full-profile sampling at high cadence, and so constitute yet stronger drivers for large solar-telescope aperture than anything advocated so far.

Although the fibril canopy seen in H$\alpha$ and partially in Ca II 8542 Å (Cauzzi et al. 2009) has less direct coronal interest, it does have the virtue of mapping closed chromospheric fields, including the long-reaching connections that make magnetic-concentration "fly-by" encounters more important in quiet-sun magnetic energy loading than flux emergence and cancellation (Meyer, Mackay & Van Ballegooijen, in preparation). The very long connections that H$\alpha$ displays in active regions and along filament channels permit excessive loading. The H$\alpha$ fibril pattern should be a valuable constraint to NLFFF extrapolations that try to predict non-potential energy budgets (Bobra et al. 2008; Wiegelmann et al. 2008), but such mapping needs higher resolution than any current H$\alpha$ monitoring achieves, including the coming GONG H$\alpha$ channels. High-resolution high-cadence H$\alpha$ mapping sorely lacks not only in Hinode but generally in chromospheric research.

A below-the-canopy H$\alpha$ phenomenon that seems ripe for decisive observation and interpretation when (if) solar activity returns is photospheric reconnection in the form of Ellerman bombs (e.g., Pariat et al. 2004; Watanabe et al. 2008).

An above-the-canopy phenomenon that may be explainable via straws/RBEs is coronal and solar-wind FIP segregation, which may have to do with multi-fluid cross-field diffusion in fibril sheaths (Judge 2008).

**Simulation.** All 3D simulations so far use the instantaneous LTE Saha equation to evaluate ionization for all species. Lower-dimension tests have established that for hydrogen, with its 10 eV jump between $n = 1$ and $n = 2$, this is no go in dynamic circumstances such as flares and acoustic shocks (Heinzel 1991; Carlsson & Stein 2002; Leenaarts et al. 2007). Helium is worse due to larger $n = 1 - n = 2$ separation. Since shocks pervade the clapotisphere, produce dynamic fibrils, and impact whatever else sits in the chromosphere, the post-shock lags ("non-equilibrium") in hydrogen ionization/recombination balancing affects any chromospheric feature. In addition, the Saha assumption fails in underestimating the electron density $N_e$ in cool regions by not accounting for photoionization of the electron-donor elements (Si, Fe, Al, Mg, Ca, Na). For example, the cool clouds in Leenaarts et al. (2010) reach $N_e \approx 10^{-6}\,N_H$ whereas $N_e$ would not sink below $N_e \approx 10^{-4}\,N_H$ (the electron-donor abundance) if photoionization was taken into account. The actual lagged-hydrogen values will not sink much below $N_e \approx 10^{-2}\,N_H$. Thus, in all 3D simulations cool-cloud electron densities have been underestimated by up to $10^4$. This lack of electrons affects the H$^-$ opacity and so underestimates radiative heating by the H$^-$ bound-free continuum.

The present 3D simulations also are amiss in not yet showing all we see in the sun. Their granulation, magnetic concentrations, and dynamic fibrils are closely akin to the observed phenomena, and their box modes mimic global $p$-mode forcing, but so far they neither contain cell-spanning fibrils nor straws/RBEs. The closest to canopy fibrils are the light-blue arches in the 6th panel of Fig. 1 of Leenaarts et al. (2007) which imply much-enhanced H$\alpha$ opacity thanks to NLTE overpopulation as large as $10^8$ (green arches in the last panel). The corresponding hydrogen ground state overpopulation in the 7th panel is as large[7] as $10^{12}$! If that display is played as a movie for the whole simulation (available via Rutten 2008) these arches are seen to exist all the time but getting kicked up severely by shocks. Their enormous overpopulations result from this relentless shocking followed by fast recombination cascades into the ground state with slow equilibrium settling via collisional Ly$\alpha$ excitation. But even these overpopulated arches have insufficient H$\alpha$ opacity to be optically thick, as many observed H$\alpha$ fibrils clearly are. Perhaps they will become less transparent in non-equilibrium 3D simulations. Perhaps these need larger resolution. Perhaps these need larger extent. And perhaps longer evo-lutionary history since Nakagawa et al. (1973) already pointed out that, in mapping chromospheric fields, "the [Hα fibril] configuration is governed apparently by evalutional [sic] consequences".

---

[7] NLTE types like me used to get excited at factors of 2, called 0.3 dex by abundance determiners. When publishing Vitas et al. (2009) we much upset a manganese abundance determiner whose 0.05 dex (12%) NLTE departure we deemed (and deem) insignificant. Here we talk 12 dex – do you hear me?

Interpretation. Let me return to solar spectral line formation. For chromospheric features the basic Eddington-Barbier recipe (Rutten 2003): "find where $\tau = 1$ accounting for NLTE ionization, take that as the formation height, and then evaluate the source function there accounting for NLTE excitation including scattering and round-about photon production, take that as the emergent intensity" fails miserably. Any paper that declares formation heights this way above $h \approx 500$ km is suspect. Any paper that derives chromospheric wave propagation from such estimates is wrong. Generally one can only say that opaque chromospheric features lie somewhere between the photosphere and the telescope, and only estimate their height more precisely near or at the limb or from tracking both Doppler and proper motion along a definite trajectory. For example, RBEs have higher-up formation further out in the Hα wings, just the reverse of Eddington-Barbier estimation. Traditional inversion of RBE profiles would err dramatically.

A much better recipe is to apply cloud-model understanding for chromospheric features. However, most if not all cloud modeling so far (review by Tziotziou 2007) assumed statistical equilibrium in the footsteps of Beckers (1964), which fails at least for hydrogen and helium because no piece of chromosphere remains unshocked sufficiently long. A much rougher but better recipe is to simply set $N_e/N_H \equiv 10^{-2}$. A more refined recipe is to perform statistical-equilibrium evaluation of hydrogen and helium populations at high temperature and to then use these values also for subsequent low-temperature phases.

## 5. Conclusion

On the ground, open-telescope technology, adaptive optics, and image restoration make large solar-telescope aperture useful and promising for chromospheric research. The ATST has entered its construction phase, a key advance in this ambitious project. In space, SDO will provide continuous monitoring of the below-the-chromosphere causes and the above-the-chromosphere effects, and IRIS will add Mg II h & k as new seeing-free chromosphere diagnostic. Numerical simulations will get more realistic. Non-equilibrium cloud modeling may power reliable inversion algorithms.

For quiet areas the physics and role of spicules-II/straws/GBEs seem the most promising topic. In active regions and filament channels my bet is on quantification, perhaps even prediction, of magnetic energy loading with NLFFF methods combining magnetogram and Hα image sequences.

Finally, distincting between "quiet" and "active" chromosphere is rather artificial. Schrijver (2010) shows that the mini-CMEs of Innes et al. (2009) fit the size distribution of normal CMEs, suggesting *"a nested set of ranges of connectivity in the magnetic field in which increasingly large and energetic events can reach higher and higher into the corona"*. The principal connectivity set of the quiet chromosphere is the magnetic network set by supergranular flows; the principal connectivity set of the active chromosphere encompasses whole active regions. These topologies differ, but they harbor the same physics producing similar effects from similar causes. The Sun's fascinating physics requires holistic analysis and explanation, particularly for the chromosphere.

*Acknowledgements.* I thank the organizers for inviting me to discuss the quiet chromosphere and for gentle coercion to write this summary. K. Reardon found vintage publications in NSO's Sunspot library and on ADS. M. Loukitcheva and A. Tlatov supported me up the Cow Trail. I also acknowledge fruitful discussions at and supported by the International Space Sciences Institute at Bern, Switzerland. My workshop participation was funded by a grant of the European Commission to the Utrecht–Oslo–Stockholm collaboration in solar physics (MEST-CT-2005-020395).